# Statistical Entropy of a Schwarzschild-anti-de Sitter Black Hole

Moorad Alexanian

*Department of Physics and Physical Oceanography*
*University of North Carolina Wilmington*
*Wilmington, NC 28403-5606*

**Abstract:** We calculate the intrinsic entropy of a Schwarzschild black hole in an asymptotically anti-de Sitter space. The statistical calculation of the entropy is based on a model for particle structure that leads to confinement. The constituents of the particle are distinguishable quasiparticles. The entropy (temperature) is less (greater) than the entropy of a Schwarzschild black hole in an asymptotically flat space. The equilibrium thermodynamic states are described by pure states, myriotic fields, and the distinguishability of the internal microstates may provide a solution to the black hole information paradox by suggesting a Bose-Einstein condensate whereby the zero mass state is a limit point (or accumulation point) of condensates on the event horizon.

## 1. Introduction

Recent interesting work has been done on the use of the anti-de Sitter/conformal field theory ($AdS / CFT$) correspondence [1] to describe the distinguishable microstates of black holes [2, 3], where finite temperature configurations in the decoupled field theory correspond to black hole configurations in $AdS$ spacetimes. Black hole microstates become perfectly distinguishable when the Holevo information achieves its maximum value, which is the Bekenstein-Hawking $S_{BH}$ entropy of the black hole [2]. An interesting issue is the unitarity inherent in the time-development of a quantum state, which implies that information is conserved. It is hoped that the distinguishability of the microstates in a black hole may solve the black hole information paradox [3] when information is irretrievably lost.

Rapidly rotating black holes can produce Bose-Einstein condensate (BEC) of extremely lightweight particles in the range $\left[10^{-20} - 10^{-10}\right] eV$, not occurring in the Standard Model, but are predicted in other theories such as Quantum Chromodynamics (QCD) axion, axionlike particles in string theory, and interesting new possibilities for Dark Matter (DM) [4]. A new method has been proposed to probe BEC of ultralight Dark Matter with the aid of gravitational waves [5].

The $AdS / CFT$ correspondence can be used to describe strong interactions [6]. Similarly, particle or quasiparticle statistics can be used to generate interparticle interactions. For instance, Bose-Einstein and Fermi-Dirac statistics correspond to attractive or repulsive interactions, respectively, between classical particles, viz., the "statistical attraction" between bosons and the "statistical repulsion" between fermions [7].

In an attempt to understand the structure of elementary particles by means of statistical mechanics, the Gibbs paradox was used to describe the "confinement" of the strongly interacting constituents of an elementary particle and define a size for a particle [8]. The statistical mechanics of truly distinguishable, strictly classical particles gives rise to a heat of evaporation per particle that is independent of the number of constituents and depends only on the volume of the particle. In the words Schrödinger [9], "what is then determined (given the temperature) is not the vapour *pressure*, but the vapour *volume*, the absolute volume of the vapour, independent of the number $n$ of particles it contains. Given this "correct" volume any amount of liquid could evaporate into it, or vice versa, without disturbing the equilibrium!"

Traditionally, quasiparticles have been used in many-body theories for systems of infinite extent. However, one very important feature of the interacting system describing a particle is that it should be confined (localized in ordinary space). Therefore, the question arises if the notion of a quasiparticle is sufficiently general to describe a localized system. The surprising result is that if the Gibbs paradox is introduced quasiparticles can describe successfully a localized system, which is herein also applied to black holes.

The model for particle structure [8] gives rise to many-body forces amongst (Fermi or Bose) constituents that are purely a manifestation of the correlations resulting from the distinguishable statistics of the quasiparticles [10]. Nonetheless, the distinguishable quasiparticles behave in many respects like an ideal Bose (not Fermi) gas [10]. A relativistic quantum field theory was developed for such a model which turns out to be nonlocal [11]. The model was subsequently used to describe a black hole as being constituted by distinguishable microstates with zero-mass constituents [12].

This paper is arranged as follows. In Sec. II, we review the thermodynamics of a black hole based on a model for particle structure that leads to confinement. The particle constituents are noninteracting, distinguishable quasiparticles giving rise to the Gibbs paradox. The model was applied to a Schwarzschild black hole in an asymptotically flat space. The thermodynamic properties include a characteristic temperature $T_{bh}$, inversely proportional to the mass of the black hole and an intrinsic entropy $S_{bh}$, proportional to the area of the event horizon. In Sec. III, we consider the black hole information paradox and suggest that physical information could be stored on the event horizon. This is accomplished by the presence of a Bose-Einstein condensate of quasiparticles on the event horizon where the zero mass state is a point of accumulation of condensates. In Sec. IV, the black hole model of Sec. II is applied to a Schwarzschild black hole in an asymptotically anti-de Sitter space. The thermodynamic properties include a characteristic temperature $T_{bh}^{AdS} = (R/r_+)T_{bh}$ and an intrinsic entropy $S_{bh}^{AdS} = (r_+/R)S_{bh}$, where $R$ is the Schwarzschild event-horizon radius and $r_+$ is the Schwarzschild anti-de Sitter event-horizon radius and $(r_+/R) \leq 1$ since $\Lambda \leq 0$. Note that $T_{bh}^{AdS} S_{bh}^{AdS} = T_{bh} S_{bh}$. Finally, Sec. V summarizes our results.

## 2. Schwarzschild Black Hole

The statistical entropy of an ideal gas of distinguishable quasiparticles (microstates) with masses $m_i$ whose number is not conserved was calculated in order to describe the interior of a Schwarzschild black hole in an asymptotically flat space [12]. The grand-canonical partition function $Z$ with chemical potentials $\mu_i = 0$ for all $i$ is given by

$$Z = \prod_i \left\{ 1 - C_i(T) \right\}^{-1}, \tag{1}$$

where the partition function $C_i(T)$ for the $i$-th type of quasiparticle with mass $m_i$ is bounded

$$C_i(T) \equiv \frac{V}{h^3} \int e^{-E_i(\mathbf{p})/kT} d\mathbf{p} = \frac{4\pi}{c^3 h^3} \left( m_i c^2 \right)^2 (kT) V K_2 \left( m_i c^2 / kT \right) \leq 1, \tag{2}$$

where

$$E_i(\mathbf{p}) = \sqrt{c^2 p^2 + m_i^2 c^4} \geq m_0 c^2 \tag{3}$$

with $m_0$ the lowest mass of the system, $V$ is the volume, and $K_\mu(x)$ is the modified Bessel function of the second kind with subscript $\mu$. If the lowest mass $m_0 = 0$, then inequality (1) implies that for fixed volume $V$, $T \leq T_{max}$ with

$$V T_{max}^3 = \pi^2 (\hbar c / k)^3 \tag{4}$$

since $(d/dx)\left[x^2 K_2(x)\right] = -x^2 K_1(x) < 0$ for $0 \leq x < \infty$ and $z^\nu K_\nu(z) \to 2^{\nu-1}\Gamma(\nu)$ as $|z| \to 0$ for $\Re\nu > 0$.

The internal energy $U$ is

$$U = -\left[\frac{\partial(\ln Z)}{\partial(1/kT)}\right]_V = \sum_i \frac{B_i(T)}{1 - C_i(T)}, \tag{5}$$

where

$$B_i(T) = \frac{V}{h^3} \int e^{-E_i(\mathbf{p})/kT} E_i(\mathbf{p}) d\mathbf{p}. \tag{6}$$

The entropy $S$ is given by

$$S = k\left[\frac{\partial(T \ln Z)}{\partial T}\right]_V = -k \sum_i \ln\left[1 - C_i(T)\right] + U/T. \tag{7}$$

The internal energy $U$ has a simple pole, for fixed $V$, at $T = T_{max}$ owing to the lowest mass state. Therefore, near the singularity we have that

$$S = \frac{U}{T} - k \ln Z \approx U/T_{max} \tag{8}$$

as $T \to T_{max}$.

Now the event horizon with proper area $A = 4\pi R^2$ encloses a finite proper volume $V = \pi^2 R^3$ and so one has for the entropy and temperature of a Schwarzschild black hole with $U = Mc^2$ we have that [12]

$$S_{bh} = \frac{1}{2\pi}\left(kc^3 A/4\hbar G\right), \tag{9}$$

and

$$T_{bh} = 4\pi\left(\hbar c^3/8\pi kGM\right), \tag{10}$$

where we have used the Schwarzschild or gravitational radius of the mass $M$, viz., $R = 2GM/c^2$. The quantities in braces in Eqs. (9) and (10) are the corresponding values obtained by Hawking [13]. The specific heat at constant volume is given by

$$C_V = \left(\frac{\partial U}{\partial T}\right)_V \approx \frac{1}{kT^2} \sum_i \frac{B_i(T)^2}{\left(1 - C_i(T)\right)^2} \approx \frac{1}{kT^2}\left(\sum_i \frac{B_i(T)}{1 - C_i(T)}\right)^2 = \frac{U^2}{kT^2}, \tag{11}$$

where in both approximations we have kept the leading singularity as $T \to T_{max}$. Condition $U/kT \gg 1$ is satisfied for Schwarzschild black holes with masses $M$ much greater than the Planck mass, $M_p = (\hbar c/G)^{1/2} \approx 2 \times 10^{-5} g$, since (10) gives that $U/kT_{bh} = Mc^2/kT_{bh} = 2(M/Mp)^2 \gg 1$. Therefore, in obtaining result (11), we have neglected the term with a simple pole and kept only the term with a pole of order 2 at $C_i(T) = 1$ coming from the $m = 0$ quasiparticles. Mass $m = 0$ is a (nonisolated) limit point (or accumulation point) of condensates (see below Sec. III).

We have then for the specific heat of a Schwarzschild black hole of mass $M$ in an asymptotically flat space

$$C_V^{bh} = k \left( \frac{2GM^2}{\hbar c} \right)^2 ,$$ (12)

where we have used (10) for the temperature of the black hole and the internal energy $U = Mc^2$.

The description of an equilibrium state as a pure state was proposed as a generalization of the formalism of axiomatic quantum field theory to systems with an infinite number of particles but finite densities [14]. This approach indicates that the notions of "the approach to equilibrium" in statistical mechanics and that of the "asymptotic condition" in axiomatic quantum field theory are the same. A Hilbert space formulation of quantum statistical mechanics was proposed [15] by using the notion of myriotic fields [16]. The pure nature of the equilibrium state together with the distinguishability of the internal microstates suggests that information may be stored inside the black hole thus possibly solving the black hole information paradox when information is presumably irretrievably lost [3].

## 3. Information Paradox

Quantum determinism follows from the future time development of a pure quantum state being determined by a unitary evolution operator acting on the state of the system. This also is associated with the reversibility whereby evolution operator has an inverse and so past states of the system are similarly uniquely determined. Accordingly, unitarity implies that information is conserved in the quantum sense. Now physical information may permanently disappear in a black hole by many different physical states collapsing into the same state. This is referred as the information paradox for black holes. The partition function $Z$ in (1) is the expected value

$$Z = \left( \Phi, e^{-\frac{1}{kT} \sum_i E_i(\mathbf{p}) n_i} \Phi \right),$$ (13)

where $\Phi$ is the product of conditioned equidistribution states, which is a functional $\phi(n)$ of the occupation function $n_i$, and is the equilibrium state of a system of noninteracting particles or quasiparticles of masses $m_i$ [15]. It is clear from inequality (2) that the zero mass quasiparticle determines the volume of the black hole. Now in order to provide a mechanism whereby information can be stored in the black hole, we consider macroscopic occupation of many zero-mass states with the $m = 0$ state a limit point (or accumulation point) of condensates. Note that the mere macroscopic occupation of the $m = 0$ state will not do since many different physical states collapsing into the same state will not solve the information paradox. However, the occurrence of macroscopic occupation of a dense set of zero-mass states can be used to store information in those states and thus resolve the information paradox. The notion of limiting point of condensates, which gives rise to spatially nonuniform condensates, was considered for one- and two-dimensional superfluids since such lower dimensional systems could not possess a uniform condensate [17].

This characterization of where the information can be stored in the black hole suggests that the condensate resides on the event horizon and so is the information content of the black hole. Hawking suggested in 2015 that the information sucked into a black hole is permanently encoded on the boundary (or event horizon) of the black hole [18].

## 4. Anti-de-Sitter Black Hole

In Sec. II, the thermodynamics of a Schwarzschild black hole in an asymptomatically flat space was obtained with the aid of constituent quasiparticles that follow strictly classical statistics and so are distinguishable and give rise to the Gibbs paradox [8]. The quasiparticles represent the internal distinguishable microstates of the black hole [12].

The Einstein equations with a negative cosmological constant $\Lambda$ admit black hole solutions which are asymptotic to anti-de Sitter space [19, 20] with metric

$$ds^2 = -\left(1 - \frac{2GM}{r} - \frac{\Lambda r^2}{3}\right)dt^2 + \left(1 - \frac{2GM}{r} - \frac{\Lambda r^2}{3}\right)^{-1} dr^2 + r^2\left(d\theta^2 + \sin^2\theta d\varphi^2\right). \quad (14)$$

The metric (14) possesses a single event horizon at $r = r_+ \leq 2GM/c^2$ since $\Lambda \leq 0$, viz., an event horizon smaller than that of a Schwarzschild black hole in an asymptotically flat space.

The results of Sec. II apply equally to anti-de Sitter black holes but with area $A = 4\pi R^2$, volume $V = \pi^2 r_+^3$ and

$$r_+ = R\left[1 + \left(\frac{\Lambda}{3}R^2\right) + 3\left(\frac{\Lambda}{3}R^2\right)^2 + 3\left(\frac{\Lambda}{3}R^2\right)^3 + \left(\frac{\Lambda}{3}R^2\right)^4 + ...\right], \quad (15)$$

with the aid of the Newton-Raphson iteration and $R = 2GM/c^2$, the Schwarzschild event-horizon radius.

One obtains for the entropy and temperature of the Schwarzschild black hole in an asymptotically anti-de Sitter space

$$S_{bh}^{AdS} = \frac{1}{2\pi}\left(\frac{r_+}{R}\right)\left(\frac{kc^3 A}{4\hbar G}\right), \quad (16)$$

and

$$T_{bh}^{AdS} = 4\pi\left(\frac{R}{r_+}\right)\left(\frac{\hbar c^3}{8\pi kGM}\right). \quad (17)$$

Note that the product $T_{bh}^{AdS} S_{bh}^{AdS}$ is independent of the cosmological constant $\Lambda$ and is equal to the product $T_{bh} S_{bh}$, which is consistent with (4). The above black holes have positive specific heat and can be in stable equilibrium with thermal radiation at a fixed temperature. One obtains that

$$C_V^{AdS} \approx \frac{U^2}{kT^2} = \left(\frac{r_+}{R}\right)^2 C_V^{bh} \le C_V^{bh} \tag{18}$$

since $\Lambda \le 0$, $U = Mc^2$, and the temperature is given by (17).

## 5. Summary and Conclusion

The thermodynamic properties of a Schwarzschild black hole in an asymptotically anti-de Sitter space follows from a model of particle physics where the constituents of the particle are noninteracting, distinguishable quasiparticles. The model was applied previously to a Schwarzschild black hole in an asymptotically flat space. The thermodynamic properties included a characteristic temperature $T_{bh}$, inversely proportional to the mass of the black hole and an intrinsic entropy $S_{bh}$, proportional to the area of the event horizon.

The thermodynamic properties for the anti-de Sitter black hole include also a characteristic temperature $T_{bh}^{AdS} = (R/r_+) T_{bh}$ and an intrinsic entropy $S_{bh}^{AdS} = (r_+/R) S_{bh}$, where $R$ is the Schwarzschild event-horizon radius and $r_+$ is the Schwarzschild anti-de Sitter event horizon radius and $(r_+/R) \le 1$ since $\Lambda \le 0$. Note that $T_{bh}^{AdS} S_{bh}^{AdS} = T_{bh} S_{bh}$.

The nonseparable nature of the Hilbert space of equilibriums states, described by myriotic fields, together with the distinguishability of the internal microstates of the equilibrium states suggests that physical information could be stored on the event horizon of a black hole. The mechanism would be the presence of a Bose-Einstein condensate whereby the zero mass state is a limit point (or accumulation point) of condensates on the event horizon.